\documentclass[aps,twocolumn,amsmath,amssymb,a4paper]{revtex4}

\usepackage[utf8]{inputenc}
\usepackage[english]{babel}
\usepackage{amsmath}
\usepackage{amsfonts}
\usepackage{amssymb}
\usepackage{amsthm}
\usepackage{graphicx}
\usepackage{color}
\usepackage{xcolor}
\usepackage{fullpage}
\usepackage{fancybox}
\usepackage{float}
\usepackage{placeins}
\usepackage{tikz}
\usetikzlibrary{trees,arrows,decorations.pathmorphing,positioning,shapes}
\usetikzlibrary{decorations.markings,fadings}
\usepackage{verbatim}

\theoremstyle{definition}

\newcommand{\beq}{\begin{equation}}
\newcommand{\eeq}{\end{equation}}

\renewcommand{\S}{\text{S}}
\renewcommand{\P}{\text{P}}

\renewcommand{\H}{\text{H}}

\renewcommand{\b}[1]{\ensuremath{\mathbf{#1}}}

\newcommand{\w}{w_{\text{ee}}}
\newcommand{\x}{\text{x}}
\newcommand{\ee}{\text{ee}}
\newcommand{\tc}{\text{c}}

\newcommand{\bra}[1]{\ensuremath{\langle #1 \vert}}
\newcommand{\ket}[1]{\ensuremath{\vert #1  \rangle}}
\renewcommand{\H}{\ensuremath{\text{H}}}
\renewcommand{\b}[1]{\ensuremath{\mathbf{#1}}}
\newcommand{\KS}{\ensuremath{\text{KS}}}

\newcommand{\lr}{\ensuremath{\text{lr}}}
\newcommand{\sr}{\ensuremath{\text{sr}}}

\newcommand{\EE}{\text{EE}}

\DeclareMathOperator{\erf}{erf}

\long\def\ignore#1{}

\begin{document}
\title{Calculating excitation energies by extrapolation along adiabatic connections
} 

\author{Elisa Rebolini$^{1,2,4}$}\email{erebolini@kjemi.uio.no}
\author{Julien Toulouse$^{1,2}$}\email{julien.toulouse@upmc.fr}
\author{Andrew M. Teale$^{3,4}$}\email{andrew.teale@nottingham.ac.uk}
\author{Trygve Helgaker$^{4}$}\email{t.u.helgaker@kjemi.uio.no}
\author{Andreas Savin$^{1,2}$}\email{savin@lct.jussieu.fr}
\affiliation{ $^1$Sorbonne Universit\'es, UPMC Univ Paris 06, UMR
  7616, Laboratoire de Chimie Th\'eorique, F-75005 Paris,
  France\\ $^2$CNRS, UMR 7616, Laboratoire de Chimie Th\'eorique,
  F-75005 Paris, France\\ 
  $^3$School of Chemistry, University of
  Nottingham, University Park, Nottingham NG7 2RD, United
  Kingdom\\ $^4$Centre for Theoretical and Computational Chemistry,
  Department of Chemistry, University of Oslo, P.O. Box 1033 Blindern,
  N-0315 Oslo, Norway
}

\date{January 30, 2015}

\begin{abstract}
  In this paper, an alternative method to range-separated
  linear-response time-dependent density-functional theory and
  perturbation theory is proposed to improve the estimation of the
  energies of a physical system from the energies of a partially
  interacting system. Starting from the analysis of the Taylor
  expansion of the energies of the partially interacting system around
  the physical system, we use an extrapolation scheme to improve the
  estimation of the energies of the physical system at an intermediate
  point of the range-separated or linear adiabatic connection where
  either the electron--electron interaction is scaled or only the
  long-range part of the Coulomb interaction is included.  The
  extrapolation scheme is first applied to the range-separated
  energies of the helium and beryllium atoms and of the hydrogen
  molecule at its equilibrium and stretched geometries.  It improves
  significantly the convergence rate of the energies toward their
  exact limit with respect to the range-separation parameter.  The
  range-separated extrapolation scheme is compared with a similar
  approach for the linear adiabatic connection, highlighting the
  relative strengths and weaknesses of each approach.

\end{abstract}

\maketitle

\section{Introduction}
The calculation of excitation energies in density-functional theory
(DFT) is nowadays mostly done by means of linear response theory in
the time-dependent framework. Linear-response time-dependent
DFT (TDDFT)~\cite{Casida1995} exhibits an
excellent price--performance ratio and is, within the usual adiabatic
semi-local approximations, very successful at describing 
excitations to low-lying valence states. However, these approximations
introduce several limitations, especially
for the treatment of static correlation~\cite{Gritsenko2000}, Rydberg
and charge-transfer excitations~\cite{Casida1998, Dreuw2003}, and
double or multiple excitations~\cite{Maitra2004}.

Time-dependent theory is, however, not mandatory for calculating
excitation energies, as stated by the Hohenberg--Kohn
theorem~\cite{Hohenberg1964}. Indeed, several time-independent DFT
approaches for calculating excitation energies exist and are currently
being developed.  These include ensemble DFT~\cite{Theophilou1979,
  Gross1988a, Pastorczak2013, Franck2013, Yang2014,
  Pribram-Jones2014}, $\Delta$SCF~\cite{Gunnarsson1976, Ziegler1977,
  Barth1979, Kowalczyk2011, Seidu2014} and related
methods~\cite{Ferre2002, Gilbert2008, Krykunov2013, Evangelista2013},
or perturbation theory~\cite{GorLev-PRB-93, Gorling1995, Gorling1996,
  Filippi1997} along the standard adiabatic connection using the
non-interacting Kohn--Sham (KS) Hamiltonian as the zero-order
Hamiltonian.

Range-separated DFT constitutes an alternative to standard KS
DFT~\cite{Hohenberg1964, Kohn1965} where the physical electronic
Hamiltonian is replaced not by an effective non-interacting
Hamiltonian but by a partially interacting Hamiltonian that
incorporates the long-range part only of the electron--electron
interaction~\cite{Savin1996, Yang1998, Pollet2003, Savin2003,
  Toulouse2004}. This partially interacting Hamiltonian corresponds to
an intermediate point along a range-separated adiabatic connection,
which links the KS Hamiltonian to the physical Hamiltonian by
progressively switching on the long-range part of the two-electron
interaction, whilst simultaneously modifying the one-electron
potential so as to maintain a constant ground-state density.

In range-separated time-dependent DFT, the excitation energies of the
long-range interacting Hamiltonian act as starting approximations to
the excitation energies of the physical system and are corrected using
a short-range density-functional kernel, in the same manner as the KS
excitation energies act as starting approximations in linear-response
TDDFT.  Several such range-separated linear-response schemes have been
developed, in which the short-range part is described by an
approximate adiabatic semi-local density-functional kernel and the
long-range linear-response part is treated at the
Hartree--Fock~\cite{Rebolini2013, Toulouse2013a, FroKneJen-JCP-13,
  Hedegard2013}, multi-configuration self-consistent field
(MCSCF)~\cite{FroKneJen-JCP-13, Hedegard2013}, second-order
polarization-propagator approximation
(SOPPA)~\cite{Hedegard2013}, or density-matrix functional
theory (DMFT)~\cite{Per-JCP-12} level.

Within the time-independent framework, a standard method for improving
upon the excitation energies of the partially interacting Hamiltonian
would be to use perturbation theory. However, given that perturbation
theory in its standard Rayleigh--Schr\"odinger based formulation does
not keep the ground-state density constant at each order in the
perturbation, it has not led to a systematic
improvement~\cite{Rebolini2015}.

In this work, we propose a time-independent alternative method for
correcting the excitation energies of the partially-interacting
system, based on extrapolation along the range-separated adiabatic
connection. Given that the long-range part of the interaction is
included in the partially interacting system, its excitation energies
constitute better approximations to the energies of the physical
system than do the excitation energies of the KS system.  The analysis
of the Taylor expansion of the energies in the range-separation
parameter $\mu$ about the physical system ($\mu \to +\infty$)
presented in Ref.~\onlinecite{Rebolini2014} shows that the energies of
the partially interacting system converge towards their physical
limits as $\mu^{-2}$. Using this information, it is possible to
develop a scheme for extrapolating the energies of the physical system
from the energies of the partially interacting system by following the
ideas of Refs.~\onlinecite{Savin2011, Savin2014}. This extrapolation
scheme involves low-order derivatives of the energies with respect
to $\mu$ and constitutes an alternative to perturbation theory and to
range-separated TDDFT~\cite{Tawada2004, Fromager2013, Rebolini2013}.

The extrapolation scheme is also applied to the linear adiabatic
connection, where the interaction is scaled by a parameter $\lambda$
going from 0 to 1, and where the analysis of the excitation energies
around $\lambda=1$ provides the required information to improve the
estimation of the energies of the physical system from an intermediate
point of the connection.

The expression for the energies of the partially interacting system and for
their extrapolations are given in Section~\ref{sec:extrapolation} 
for the range-separated and linear adiabatic connections. The extrapolation is
subsequently applied to the range-separated energies 
of the helium and beryllium atoms and
of the hydrogen molecule at its equilibrium and stretched geometries;
for helium, we also use the linear adiabatic connection. The
computational details are given in Section~\ref{sec:compdet} and the
results are discussed in Section~\ref{sec:chap5results}.

\section{Energy extrapolation along the range-separated and linear 
  adiabatic connections}
\label{sec:extrapolation}
\subsection{Range-separated adiabatic connection}
Range-separated DFT uses a partially interacting system, where the
long-range part of the Coulomb interaction is included instead of the
more traditional non-interacting KS system---see, for example,
Ref.~\onlinecite{Toulouse2004}.  In terms of the long-range (lr)
electron--electron interaction operator
\begin{eqnarray}
  \hat{W}_\text{ee}^{\lr,\mu} \!=\! \frac{1}{2} \iint
  w_\text{ee}^{\lr,\mu}(r_{12}) \hat{n}_2(\b{r}_1,\b{r}_2) \mathrm
  d\b{r}_1 \mathrm d\b{r}_2,
\end{eqnarray}
where $\hat{n}_2(\b{r}_1,\b{r}_2)$ is the pair-density operator and
$w_\text{ee}^{\lr,\mu}(r_{12})$ is the error-function interaction
\begin{equation}
  w_\text{ee}^{\lr,\mu}(r_{12}) = \frac{\erf(\mu r_{12})}{r_{12}},
\end{equation}
the Hamiltonian of the partially interacting system is given by
\begin{equation}
  \hat{H}^{\lr,\mu} = \hat{T} + \hat{V}_\text{ne} +
  \hat{W}^{\lr,\mu}_\text{ee} + \hat{\bar{V}}_{\H
    \text{xc}}^{\sr,\mu}.
\label{Hmu}
\end{equation}
The parameter $\mu$ controls the range of the separation, with $1/\mu$
acting as a smooth cut-off radius.  This Hamiltonian also contains the
short-range Hartree--exchange--correlation potential operator
$\hat{\bar{V}}_{\H \text{xc}}^{\sr,\mu}$, whose role is to  ensure that the
ground-state density of the partially-interacting system 
\begin{equation}
n_0(\b{r}) = \bra{\Psi_0^{\mu}} \hat{n}(\b{r}) \ket{\Psi_0^{\mu}},
\end{equation}
is equal to the
ground-state density of the physical system for all $\mu$. Here
$\Psi_0^{\mu}$ is the ground-state wave function of the partially
interacting Hamiltonian and $\hat{n}(\b{r})$ is the density operator.
The remaining terms in the Hamiltonian of Eq.~\eqref{Hmu} are the usual
kinetic-energy operator $\hat{T}$ and nuclear--electron interaction
operator $\hat{V}_\text{ne} = \int v_\text{ne}(\b{r}) \hat{n}(\b{r})
\mathrm d\b{r}$. 

The eigenvectors and eigenvalues of $\hat{H}^{\lr,\mu}$ are the
ground- and excited-state wave functions $ | \Psi_k^{\mu} \rangle$ and
energies ${\cal E}_k^{\mu}$ of the partially interacting system
\begin{equation}
  \hat{H}^{\lr,\mu} | \Psi_k^{\mu} \rangle = {\cal E}_k^{\mu} |
  \Psi_k^{\mu} \rangle.
\label{HlrmuPsi}
\end{equation}
These excited-state wave functions and energies provide natural first
approximations to the excited-state wave functions and energies of the
physical system. For $\mu=0$, they reduce to the single-determinant
eigenstates and associated energies of the non-interacting KS
Hamiltonian,
\begin{equation}
  \hat{H}^{\KS} | \Phi_k^{\KS} \rangle = {\cal E}_k^{\KS} |
  \Phi_k^{\KS} \rangle,
\end{equation}
while, for $\mu \to \infty$, they reduce to the excited-state wave
functions and energies of the physical Hamiltonian
\begin{equation}
\hat{H} | \Psi_k \rangle = E_k | \Psi_k \rangle.
\end{equation}

In Ref.~\onlinecite{Rebolini2014}, it was shown that the asymptotic
expansion of the total energy of state $k$ around the physical
system is
\begin{equation}
  {\cal E}_k^{\mu}= E_k + \dfrac{1}{\mu^2} E_k^{(-2)} +
  \dfrac{1}{\mu^3} E_k^{(-3)} + \mathcal{O}\left(\dfrac{1}{\mu^4}
  \right),
  \label{eq:E0infty}
\end{equation}
where $E_k^{(-2)}$ and $E_k^{(-3)}$ are the corrections entering at
the second and third powers of $1/\mu$, respectively.  Following the
scheme proposed in Refs.\;\onlinecite{Savin2011,Savin2014}, it is possible
to estimate the energy of the physical system $E_k$ from the energy of
the partially interacting system ${\cal E}_k$ and its first- and
second-order derivatives with respect to $\mu$.

>From the Taylor expansion of the energies when $\mu \to \infty$, 
the first-order derivatives of the energies with respect to $\mu$
behave as
\begin{equation}
  \dfrac{\partial {\cal E}_k^{\mu}}{\partial \mu} =
  -\dfrac{2}{\mu^3}E_k^{(-2)} + \mathcal{O} \left( \dfrac{1}{\mu^4}
  \right),
\end{equation}
around the real system.  Inserting this into Eq.~\eqref{eq:E0infty},
the exact energies $E_k$ can be written as a function of the energies
along the adiabatic connection and of their first-order derivative as
\begin{equation}
  E_k = {\cal E}_k^{\mu} + \dfrac{\mu}{2} 
  \dfrac{\partial  {\cal E}_k^{\mu}}{\partial \mu} 
  + \mathcal{O}\left(\dfrac{1}{\mu^3}\right).
\end{equation}
This scheme gives extrapolated energies 
\begin{equation}
  E^{\EE,\mu}_k={\cal E}_k^{\mu} + \dfrac{\mu}{2} \dfrac{\partial
    {\cal E}_k^{\mu}}{\partial \mu},
\label{eq:extrapolation}
\end{equation}
that are correct up to and including the second power of $1/\mu$
relative to the energies of the physical system.  The correction given
by the extrapolation scheme vanishes at $\mu=0$ by construction, but
should improve the description of the energies as soon as the
interaction is switched on. One should note that the absence of
a correction at $\mu=0$ is only due to the choice of $1/\mu^k$ as the basis
for the expansion. Other basis functions such as $\mu^2/(a + \mu^5)$
would lead to a correction at $\mu=0$ but are not considered in this
work.

A more elaborate scheme can be developed by using also the correction
$E_k^{(-3)}$ and the second-order derivative.  In this case, the
first- and second-order derivatives are given by
\begin{align}
  \dfrac{\partial {\cal E}_k^{\mu}}{\partial \mu} &=
  -\dfrac{2}{\mu^3}E_k^{(-2)} -\dfrac{3}{\mu^4}E_k^{(-3)} 
+  \mathcal{O} \left( \dfrac{1}{\mu^5} \right)
, \\
  \dfrac{\partial^2 {\cal E}_k^{\mu}}{\partial \mu^2} &=
  \dfrac{6}{\mu^4}E_k^{(-2)} + \dfrac{12}{\mu^5}E_k^{(-3)}
 +  \mathcal{O} \left( \dfrac{1}{\mu^6} \right)
,
\end{align}
and, after eliminating $E_k^{(-2)}$ and $E_k^{(-3)}$, the extrapolated
energies become
\begin{equation}
  E^{\EE 2,\mu}_k={\cal E}_k^{\mu} + \mu \dfrac{\partial {\cal
      E}_k^{\mu}}{\partial \mu} + \dfrac{\mu^2}{6} \dfrac{\partial^2
    {\cal E}_k^{\mu}}{\partial \mu^2}.
\label{eq:extrapolation2}
\end{equation}
Higher-order derivatives should further reduce errors.  Additionally,
several points along the adiabatic connection could be used to perform
the extrapolation to increase the accuracy of the extrapolated
energies. However, only first- and second-order corrections at a
single point of the adiabatic connection are considered hereinafter.

\subsection{Linear adiabatic connection}
If the linear adiabatic connection is performed, then the partially
interacting Hamiltonian is defined as $\hat{H}^{\lambda} = \hat{T} +
\lambda \hat{W}_{\ee} + \hat{V}^{\lambda}$ where $\hat{V}^{\lambda}$
is adjusted to keep the ground-state density constant.  This potential
can be expressed in terms of the connecting parameter $\lambda$ as
\begin{equation}
  \hat{V}^{\lambda} = \hat{V}_\text{ne} + (1-\lambda)\hat{V}_{\H \x} +
  \hat{V}_\tc - \hat{V}_\tc^\lambda,
\end{equation}
where $\hat{V}_\tc^\lambda$ enters at second order in $\lambda$ and is
equal to $\hat{V}_\tc$ at $\lambda=1$.  The energies of the
partially interacting system can then be expanded around the physical
system as
\begin{equation}
  {\cal E}^{\lambda}_k = E_k + (\lambda-1)E_k^{(1)} +
  (\lambda-1)^2E_k^{(2)} + \mathcal{O}(\lambda-1)^3,
\end{equation}
where $E_k^{(1)}$ and $E_k^{(2)}$ are the contributions entering at
the first and second power of $(\lambda-1)$, respectively. 
As in the range-separated case, by differentiation with respect to
$\lambda$, it is then possible to extrapolate the energies of the
physical system at first order by considering only the correction
$E_k^{(1)}$ as
\begin{equation}
  E_k^{\EE,\lambda} = {\cal E}^{\lambda}_k+ (1-\lambda)\dfrac{\partial {\cal
      E}_k^{\lambda}}{\partial \lambda}.
  \label{eq:extrapolation lambda}
\end{equation}
When $\lambda=0$, this extrapolation is equivalent to
the first-order correction of G\"orling--Levy perturbation
theory~\cite{Gorling1995,Filippi1997}.

A second-order correction can be obtained by using also the correction
$E_k^{(2)}$. The first- and second-order derivatives are 
\begin{align}
  \dfrac{\partial {\cal E}_k^{\lambda}}{\partial \lambda} &=
  E_k^{(1)} + 2 (\lambda_1) E_k^{(2)} + \mathcal{O}(\lambda-1)^2, \\ 
  \dfrac{\partial^2 {\cal E}_k^{\lambda}}{\partial
    \lambda^2} &= 2 E_k^{(2)} + \mathcal{O}(\lambda -1),
\end{align}
 and the extrapolated energies become 
\begin{equation}
  E^{\EE 2,\lambda}_k={\cal E}_k^{\lambda} + (1- \lambda)
  \dfrac{\partial {\cal E}_k^{\lambda}}{\partial \lambda} +
  \dfrac{1}{2} (1-\lambda)^2 \dfrac{\partial^2 {\cal
      E}_k^{\lambda}}{\partial \lambda^2}.
\label{eq:extrapolation lambda2}
\end{equation}

\section{Computational Details}
\label{sec:compdet}
Calculations were performed for the He and Be atoms and the H$_2$
molecule with a development version of the DALTON
program~\cite{Dal-PROG-11}, see
Refs.~\onlinecite{TeaCorHel-JCP-09,TeaCorHel-JCP-10,TeaCorHel-JCP-10b}.
Following the procedure of Ref.~\onlinecite{Rebolini2014}, a
full CI (FCI) calculation was first carried out to get the exact
ground-state density within the basis set considered:
uncontracted t-aug-cc-pV5Z for He,
uncontracted d-aug-cc-pVDZ for Be, and uncontracted d-aug-cc-pVTZ for H$_2$.
A Lieb optimization of the short-range potential $v^{\sr,\mu}(\b{r})$ was then
performed to reproduce the FCI density with the long-range
electron--electron interaction $\w^{\lr,\mu}(r_{12})$. Finally, an FCI
calculation was carried out with the partially-interacting Hamiltonian
constructed from $\w^{\lr,\mu}(r_{12})$ and $v^{\sr,\mu}(\b{r})$ to
obtain the zeroth-order energies and wave functions.

Starting from the analytical form of the fit given in the
supplementary material of Ref.~\onlinecite{Rebolini2014}, it is then
straightforward to calculate the analytical derivatives of the
energies with respect to $\mu$. In the linear case, a cubic fit of the
energies was performed.  The extrapolated energies were calculated
using
Eqs.~\eqref{eq:extrapolation},~\eqref{eq:extrapolation2},~\eqref{eq:extrapolation
  lambda}, and~\eqref{eq:extrapolation lambda2}.

All the unextrapolated curves shown hereinafter correspond to the
curves of Ref.~\onlinecite{Rebolini2014}.

\section{Results and discussion}
\label{sec:chap5results}

\subsection{Range-separated adiabatic connection of the helium atom}
\begin{figure}
  \centering
  \includegraphics[scale=1]{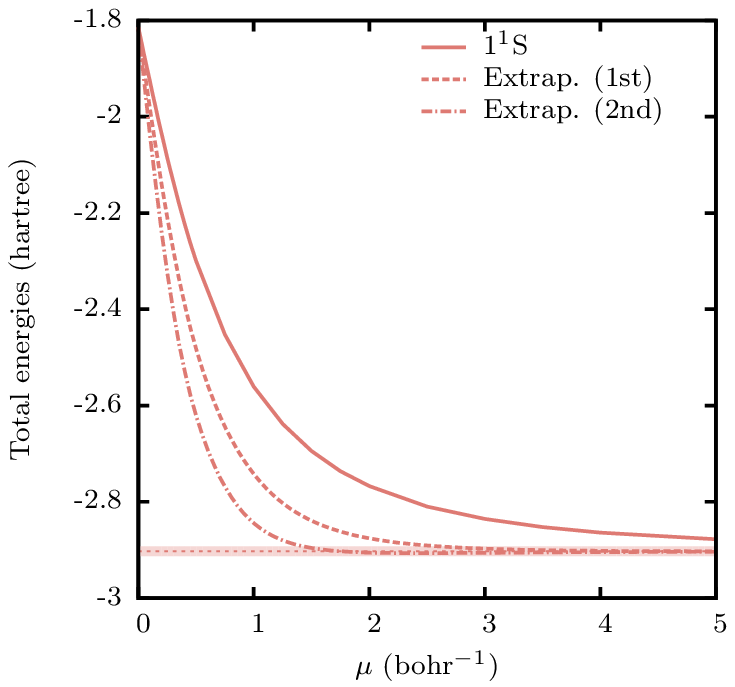}
  \includegraphics[scale=1]{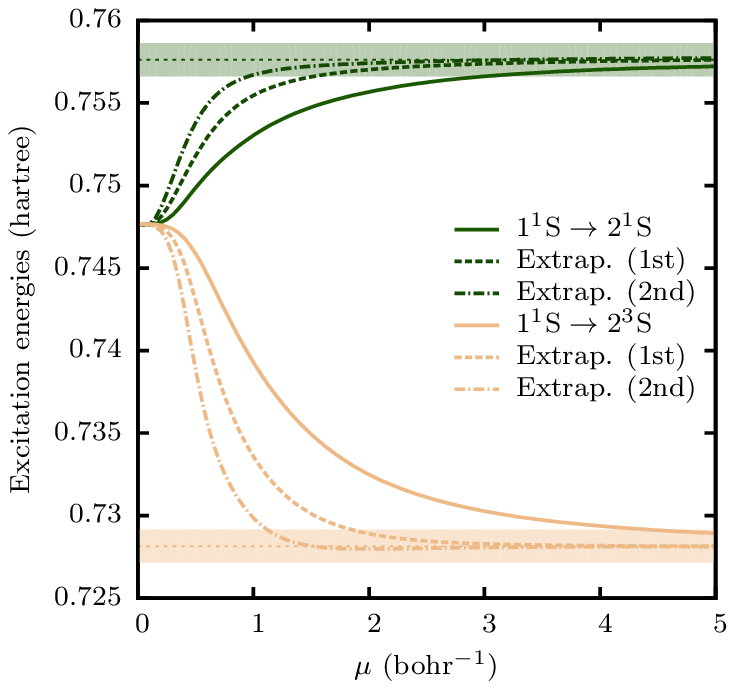}
  \includegraphics[scale=1]{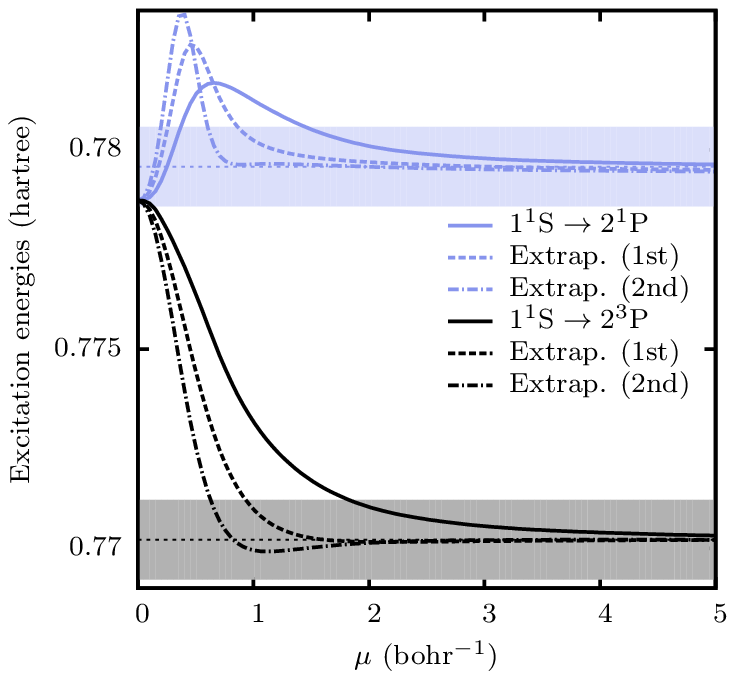}
  \caption{(Color online) Helium ground-state energy (top) ${\cal E}_0^{\mu}$, and
    first $\S$ (middle) and $\P$ (bottom) excitation energies $\Delta
    {\cal E}_k^{\mu} = {\cal E}_k^{\mu} - {\cal E}_0^{\mu}$ calculated
    without extrapolation (full lines), with first-order extrapolation
    (dashed) and second-order extrapolation (dot--dashed) as a
    function of $\mu$. The dotted horizontal lines are the physical
    energies. The colored regions represent errors of $\pm 10$ and
    $\pm 1$~mhartree for the ground-state and excitation energies,
    respectively.
    \label{fig:he_tacpv5z_svd07_0_fit}
  }
\end{figure}

\subsubsection{Ground-state energy}
The results of the first- and second-order extrapolation schemes on
the ground-state total energy of the helium atom are shown in
Figure\;\ref{fig:he_tacpv5z_svd07_0_fit} (top). By construction, the
extrapolation has no effect at $\mu=0$ and the ground-state energy of
the KS system is, therefore, unaffected by the extrapolation. However,
for $\mu > 0$, the extrapolated energies show a systematic improvement
with respect to the unextrapolated ground-state energy, that is, the
ground-state energy of the partially interacting Hamiltonian without
any correction.

Without extrapolation, a range-separation parameter of about
6\;bohr$^{-1}$ is needed to give an error smaller than
10\;mhartree relative to the energy of the physical system.  With
the first- and second-order corrections added, the same accuracy is
achieved with a range-separation parameter of only 2.8 and
1.5 bohr$^{-1}$, respectively.  

\subsubsection{Rydberg excitation energies}
Figure\;\ref{fig:he_tacpv5z_svd07_0_fit} also shows the effects of
extrapolation on the lowest Rydberg S (middle) and P (bottom)
excitation energies of helium.  The convergence of the excitation
energies towards their physical limit is overall improved by the
first- and second-order corrections with respect to the unextrapolated
curves. In fact, the range-separation parameter required to achieve an
accuracy of 1~mhartree is divided by approximately a factor 2 or a
factor 3 by the first- and second-order schemes, respectively.
For the excitation energies considered here, a range-separation value
of 2 and 1\;bohr$^{-1}$ suffices to reduce the error to less than
1\;mhartree with the first- and second-order schemes,
respectively.

The $^1\S$ and $^3\S$ excitation energies change monotonically with
increasing $\mu$. Accordingly, extrapolation provides a systematic
improvement, the sign of the derivative pulling the excitation
energies towards their physical limits at both first- and second-order
levels.

The $^3\P$ excitation energy also changes monotonically with $\mu$ and
the first-order extrapolation provides therefore a systematic
improvement. However, the first-order extrapolated energy does not
converge monotonically towards its physical limit (not visible),
leading to a slight degradation of the excitation energies around
$\mu=1.5$ bohr$^{-1}$ at second order.

Finally, the $^1\P$ excitation energy shows a non-monotonic behavior
even before extrapolation, exhibiting a ``bump'' for small $\mu$ that
is probably a basis-set effect~\cite{Rebolini2014}.  In fact, for such
small $\mu$, only the very long-range part of the interaction is
modified, which is poorly described by Gaussian basis functions. The
higher a given state is in energy, the more sensitive it becomes to
this basis-set defect.

As a consequence, the $^1\P$ excitation energy approaches its physical
limit from above, its first-order derivative changing sign around
0.7\;bohr$^{-1}$.  In this region, the extrapolated energies become
less accurate than the unextrapolated energy.  However, this behavior
is observed only in a small region.  As soon as the excitation energy
recovers a monotonic convergence towards its physical limit (for $\mu$
larger than 0.7\;bohr$^{-1}$), the energy is improved by the
extrapolation and converges faster to its physical limit.

\subsection[Range-separated adiabatic connection for the valence of Be]
           {Range-separated adiabatic connection for the valence
             excitation of the beryllium atom}

\begin{figure}
  \centering
  \includegraphics[scale=1]{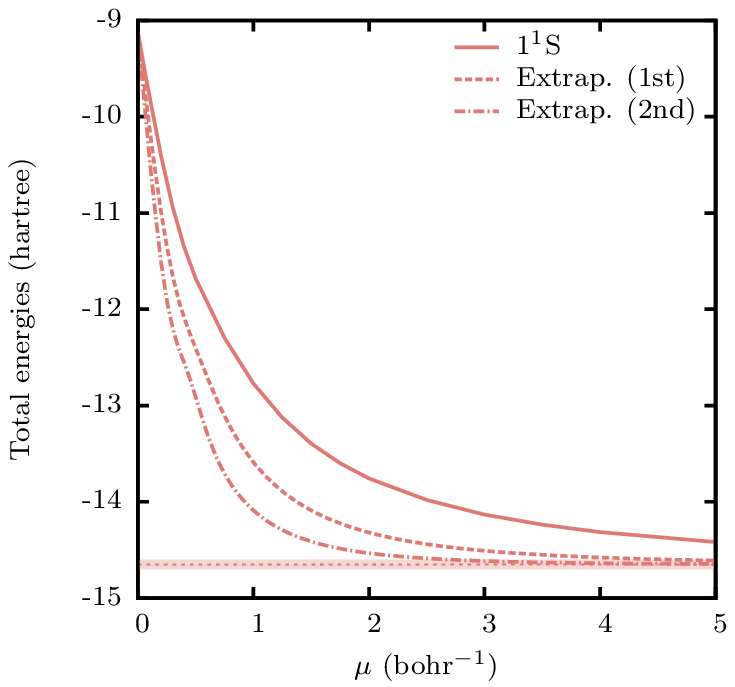}
  \includegraphics[scale=1]{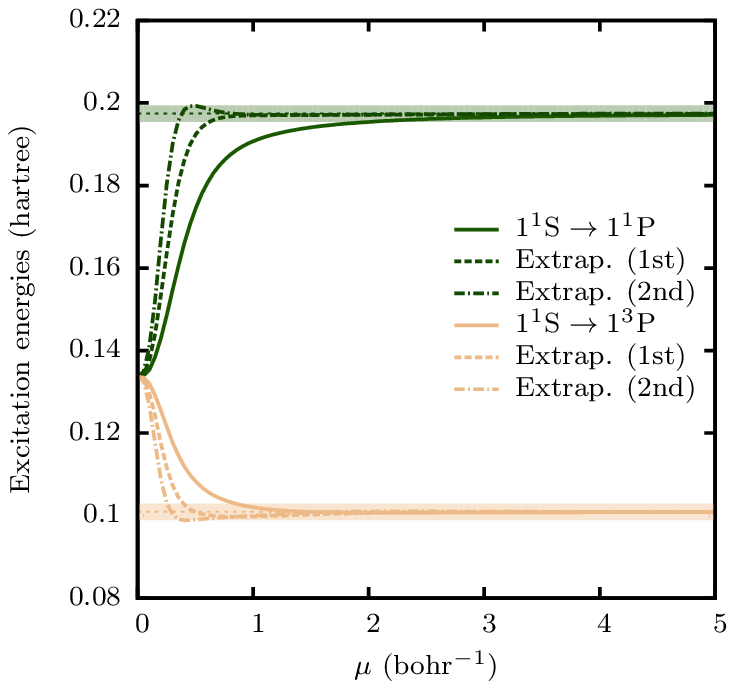}
  \caption{(Color online) Ground-state energy ${\cal E}_0^{\mu}$ (top) and excitation
    energies $\Delta {\cal E}_k^{\mu} = {\cal E}_k^{\mu} - {\cal
      E}_0^{\mu}$ (bottom) of beryllium as a function of $\mu$.  The
    unextrapolated energies are shown as full lines, the first-order
    extrapolated energies are plotted in dashed lines and the
    second-order ones in dot-dashed lines. The energies of the
    physical system are given as horizontal dotted lines. An error of
    $\pm 50$ mhartree is colored around the physical ground-state
    energy and an error of $\pm 2$ mhartree is colored around the
    physical excitation energies.
    \label{fig:be_dacpvdz_0_fit}
  }
\end{figure}

The ground-state energy of the beryllium atom is shown in
Figure~\ref{fig:be_dacpvdz_0_fit} (top).  Since beryllium has a core
orbital, the convergence of its total energies is slower than for
helium as the density is more contracted and a larger range-separation
parameter is needed to describe correctly the core region.  However,
this affects all valence states in a similar fashion (not shown here).

Extrapolation systematically improves the convergence of the
ground-state energy along the adiabatic connection.  First-order
extrapolation reduces the error to less than 50\;mhartree with
$\mu \approx 5$\;bohr$^{-1}$, an order of magnitude smaller than the
error without the extrapolation correction but still
large. Second-order extrapolation gives the same error reduction
already with $\mu \approx 3$\;bohr$^{-1}$.

The effect of the extrapolation on the valence excitation energies of
beryllium is shown in Figure\;\ref{fig:be_dacpvdz_0_fit} (bottom).  As
the errors associated with the core largely cancel in the excitation
energies, the unextrapolated excitation energies already converge
faster than do the total energies.  With first-order extrapolation, an
error smaller than 2\;mhartree is reached with $\mu \approx
0.5$\;bohr$^{-1}$, to be compared with a much larger error of
4\;hartree in the total energies with the same $\mu$ value.  The
second-order extrapolation allows one to reach the same accuracy with
a range-separation parameter as small as 0.3\;bohr$^{-1}$. However,
once again, some bumps are observed in the extrapolated energies probably due
to the limited size of the basis set. This fast convergence of the
excitation energies with respect to the range-separation parameter is
due to the fact that in beryllium, static correlation is important and
the multi-configurational character of the wave function is quickly
established when the interaction is switched on; see
Ref.~\onlinecite{Fromager2007}.

\subsection[Range-separated adiabatic connection for H$_2$] 
           {Range-separated adiabatic connection for the hydrogen
             molecule}
\begin{figure}
  \centering
  \includegraphics[scale=1]{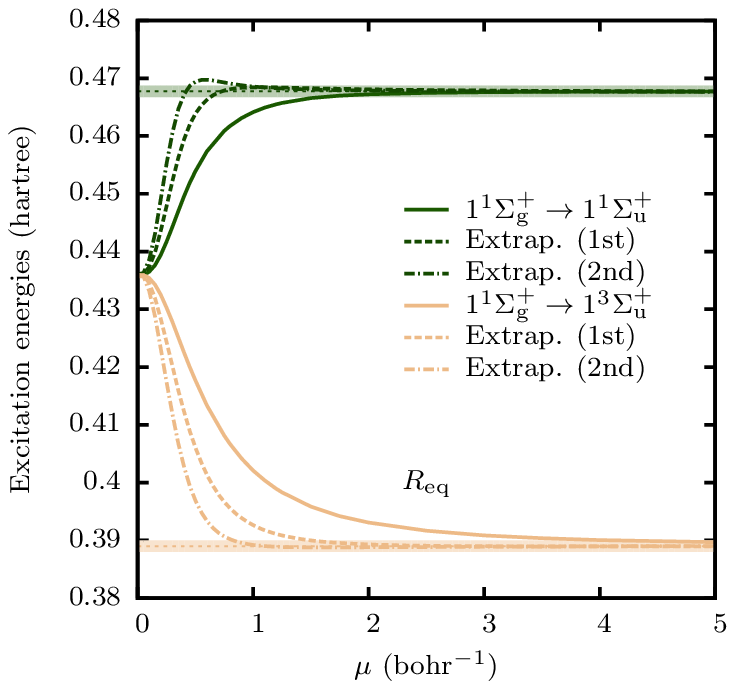}
  \includegraphics[scale=1]{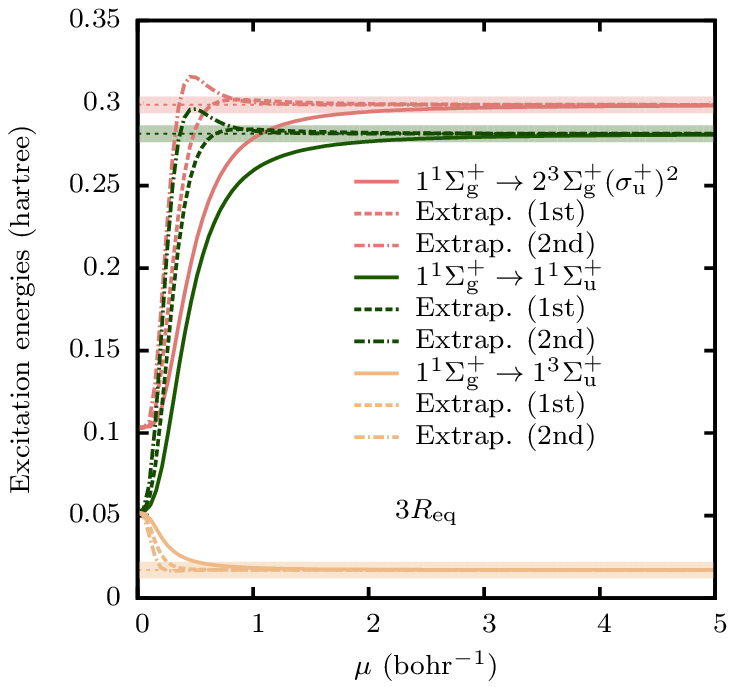}
  \caption{(Color online) Unextrapolated (full lines), first-order extrapolated
    (dashed lines) and second-order extrapolated (dot-dashed lines)
    excitation energies $\Delta {\cal E}_k^{\mu} = {\cal E}_k^{\mu} -
    {\cal E}_0^{\mu}$ of the H$_2$ molecule at the equilibrium
    internuclear distance $R_{\text{eq}}$ (top) and three times the equilibrium
    distance (bottom) as a function of $\mu$.  The excitation energies
    of the physical system $\Delta E_k = \Delta {\cal
      E}_k^{\mu\to\infty}$ are plotted as horizontal dotted lines. An
    error of $\pm 1$ mhartree is colored around the physical
    excitation energies at equilibrium and an error of $\pm 5$
    mhartree at stretched geometry.
    \label{fig:h2_dacpvtz_exc_0_fit}
  }
\end{figure}
Finally, we consider extrapolation of the lowest excitation energies
of the hydrogen molecule along the range-separated adiabatic
connection, at the equilibrium geometry and at a stretched geometry.
The results of the first- and second-order extrapolations on the
singlet and triplet $\Sigma_\text{g}^+ \to \Sigma_\text{u}^+$
excitation energies at the equilibrium geometry are shown in
Figure~\ref{fig:h2_dacpvtz_exc_0_fit} (top).  First- and
second-order extrapolations provide a systematic improvement in the
excitation energies, $\mu \approx 2$\;bohr$^{-1}$ for first order
and $\mu \approx 1$\;bohr$^{-1}$ for second order being sufficient
to reproduce the physical energies to within 1\;mhartree.

Having stretched the hydrogen molecule to three times the equilibrium
distance, we apply extrapolation to the singlet and triplet
excitations to the $1\Sigma_\text{u}^+$ state and to the double
excitation to the $2\Sigma_\text{g}^+$ state---see the bottom part of
Figure~\ref{fig:h2_dacpvtz_exc_0_fit}.  Again, the improvement is
systematic. The triplet extrapolated energy shows a monotonic behavior
with respect to $\mu$, whereas the singlet energy shows a slight bump
at 0.8\;bohr$^{-1}$.  However, all extrapolated excitation energies
converge faster than their unextrapolated counterparts.  Extrapolation
works remarkably well, reducing errors to less than 5\;mhartree
with $\mu \approx 0.6$\;bohr$^{-1}$, compared with 2\;bohr$^{-1}$
without extrapolation. In particular, extrapolation allows us to
describe double and single excitation energies equally well.  In this
case, one should note that the second-order scheme does not improve
significantly the convergence of the $1 ^1\Sigma_\text{g}^+ \to 2
^3\Sigma_\text{g}^+ (\sigma_\text{u}^+)^2 $ and $1 ^1\Sigma_\text{g}^+
\to 1 ^1\Sigma_\text{u}^+ $ excitation energies because of their
nonmonotonicity probably due to the limited basis set.

\begin{figure}
  \centering
  \includegraphics[scale=1]{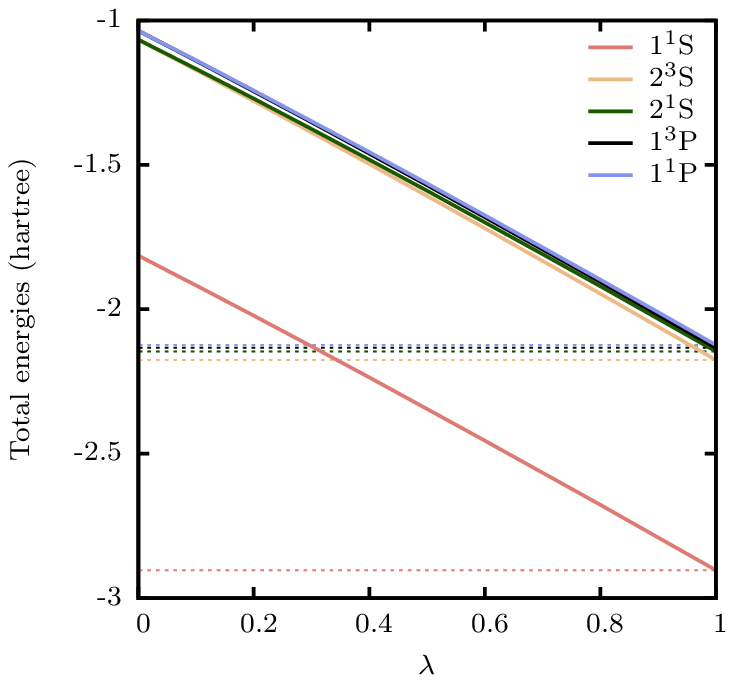}
  \includegraphics[scale=1]{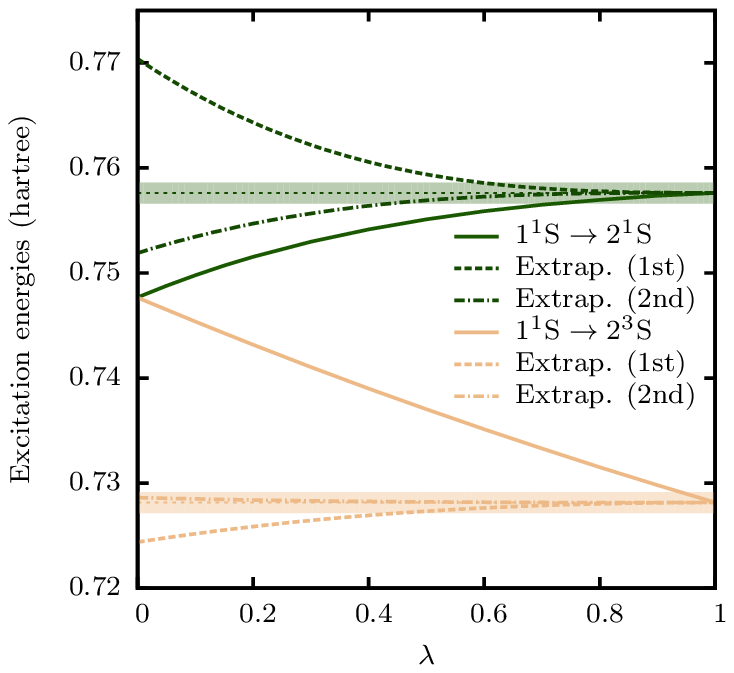}
  \includegraphics[scale=1]{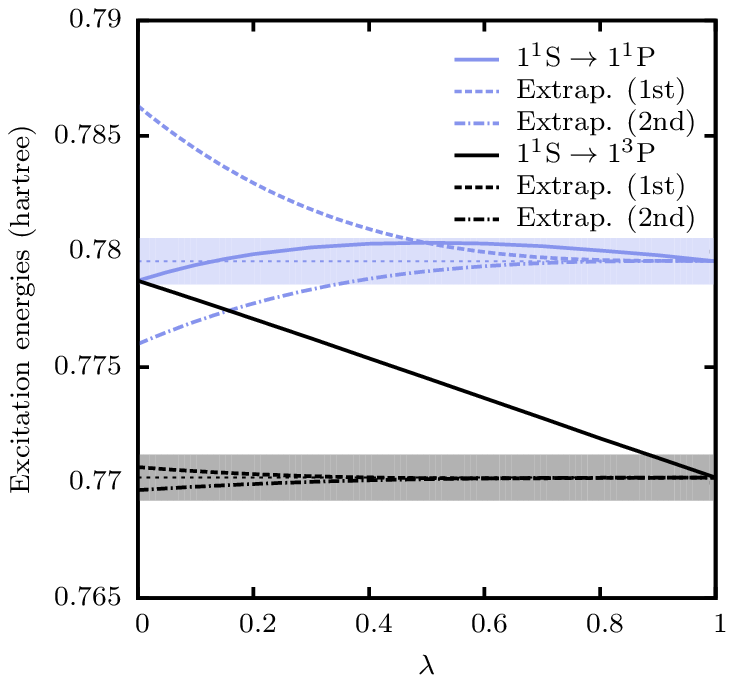}
  \caption{ (Color online) Helium ground- and excited-state energy
    (top), $\S$ excitation energies (middle), and $\P$ excitation
    energies (bottom) calculated without extrapolation (full lines),
    with first-order extrapolation (dashed), and second-order
    extrapolation (dot--dashed) along the linear connection. The
    dotted horizontal lines are the physical energies. The colored
    regions represent errors of $\pm 1$~mhartree for the
    excitation energies.
    \label{fig:he_tacpv5z_svd07_0_lambda}
  }
\end{figure}
\subsection{Linear adiabatic connection for the helium atom}
\ignore{In order to apply the extrapolation scheme on the linear
  adiabatic connection, the zero-order total and excitation energies
  of the helium atom along this connection are calculated following
  the procedure described in Chapter~\ref{chap:zeroth AC} and are
  shown in Figure~\ref{fig:he_tacpv5z_svd07_0_lambda}.}

\subsubsection{Total energies}

The total energies of the helium atom along the linear adiabatic
connection are plotted in Figure~\ref{fig:he_tacpv5z_svd07_0_lambda}
(top).  When $\lambda=0$, no interaction is included, so the KS
energies are recovered as for $\mu=0$. When $\lambda =1$, the full
interaction is present and the energies of the physical system are
recovered, which corresponds to the limit $\mu \to\infty$.  The two
limiting cases are, therefore, identical for the two adiabatic
connections but the way they are connected differs.

The evolution of the total energies with respect to $\lambda$ is
almost linear. Although this behavior is easier to predict and should
provide an efficient framework for extrapolations, the value of
$\lambda$ required to have an error of 10 mhartree is very close
to 1; in the range-separation case, an intermediate value of $\mu$ is
sufficient. In general, we note that, in the linear case, the
calculation of the wave function is by and large equally expensive at
all points along the connection since the electron-electron cusp must
always be described (except at $\lambda =0$); in the range-separated
case, by contrast, the calculation becomes less expensive with
decreasing $\mu$ since the description of the cusp is then avoided.

\subsubsection{Excitation energies and comparison with the range-separated case}
The lowest (Rydberg) excitation energies of helium along the linear
adiabatic connection are also plotted in
Figure~\ref{fig:he_tacpv5z_svd07_0_lambda} (middle and bottom). As for
the total energies, the end points are the same as in the
range-separated case but the behavior of the energies along the
connection is more linear. As in the range-separated case, the $^1\P$
excitation energy does not evolve monotonically with $\lambda$,
probably because of basis-set limitations.

When the first- and second-order extrapolation corrections are added,
a systematic improvement is observed for the triplet excitation
energies. With the second-order correction, the physical energies are
already reproduced to 1\;mhartree at
$\lambda=0$.  The singlet excitation energies are less affected by the
correction but are still overall improved, the amount of interaction
required to reproduce the physical limit within an accuracy of
1\;mhartree dropping to 50\%. Moreover, unlike in the range-separated
case, the KS excitation energy also benefits from this correction,
which no longer vanishes in this limit. Indeed, at $\lambda=0$, the
extrapolated excitation energy matches the results obtained in
Ref.~\onlinecite{Filippi1997}, using first-order G\"orling--Levy
perturbation theory.

\begin{figure}
  \centering
  \includegraphics[scale=1]{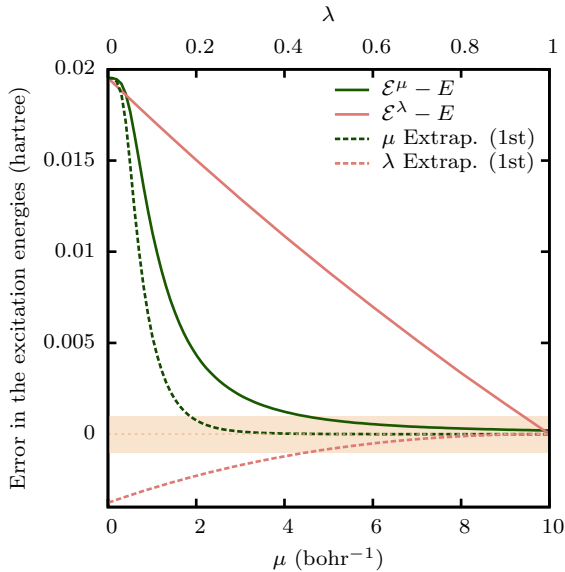}
  \caption[Comparison of the range-separated and linear adiabatic
    connections]{Error in the $^3\S$ excitation energy along the
    range-separated and linear adiabatic connections for the helium
    atom as functions of $\mu$ and $\lambda$. The unextrapolated
    energies are given in full lines and the extrapolated ones in
    dashed lines.  An error smaller than 1\;mhartree around the
    physical limit is given by the colored
    region.\label{fig:he_comp_mulambda_fit}}
\end{figure}

To compare the range-separated and linear cases, the effects of
first-order extrapolation on the lowest excitation energy of helium
are shown in Figure\;\ref{fig:he_comp_mulambda_fit} in both cases.
Without extrapolation, the scaling parameter $\lambda$ must be greater
than 0.95 to reproduce energies within 1\;mhartree. By contrast, a
small change in $\mu$ near zero gives a large change in the energies;
thus, a range-separation parameter of 2\;bohr$^{-1}$ is sufficient to
ensure the same accuracy. Clearly, the range-separated connection
includes the most significant region of the interaction first, whereas
the linear connection treats all ranges equally, independently of
their importance for the excitation energies.

For the KS system, it is obviously better to use the correction
obtained from the linear connection as the corresponding
range-separated correction vanishes.  For a partially interacting
system, the comparison is more difficult.

\section{Conclusion}

In this work, we have exploited the asymptotic behavior of the
energies of a partially interacting system along the range-separated
adiabatic connection to design an energy correction that allows us to
extrapolate to the physical energies of the system from its partially
interacting energies.  The simplest possible extrapolations were
obtained by using either only the first-order derivative of the
energies with respect to the range-separation parameter at a given
point or by using the first- and second-order derivatives.

This extrapolation scheme was tested at the FCI level of theory on the
helium and beryllium atoms and on the hydrogen molecule (at equilibrium
and at stretched geometry), where it significantly improves the
convergence of energies and excitation energies towards their physical
limits.  Moreover, the improvements are systematic, except at $\mu=0$
(where the correction is zero by construction) and in a few cases where
the partially interacting energies present a bump for small $\mu$.
In all cases, with respect to the unextrapolated case, the
extrapolation schemes reduce the smallest value of the range-separation
parameter required to reproduce the physical energies of the system
with a given accuracy by approximately a factor of 2 with the
first-order scheme and by a factor of 3 with the second-order scheme.
This is of particular relevance for truncated wave
functions, as the smaller the range-separation parameter is, the fewer
Slater determinants are needed to describe the wave functions with an
equivalent accuracy.

Finally, the extrapolation scheme was applied along the linear
adiabatic connection, where it also improves significantly the
description of the excitation energies along the connection.

All results discussed here were obtained without the use of
approximate functionals. The proposed extrapolation scheme should now
be tested in a more pragmatic case, where the potential is not
obtained by Lieb optimization but from different approximations such
as the (semi)local approximations or more interestingly approximations
where the long-range behavior of the potential is correct, such as the
optimized-effective potential (OEP)~\cite{Sharp1953,Talman1976}. The
effects of the inclusion of higher-order derivatives and of multiple
points on this extrapolation should also be explored.

\section{Acknowledgments}
 This work was supported by the Norwegian Research Council through the
 CoE Centre for Theoretical and Computational Chemistry (CTCC) Grants
 No.\ 179568/V30 and No.\ 171185/V30 and through the
 European Research Council under the European Union Seventh Framework
 Program through the Advanced Grant ABACUS, ERC Grant Agreement
 No.\ 267683. A. M. T. is grateful for support from the Royal Society
 University Research Fellowship scheme.

\end{document}